\documentclass[
aps,%
11pt,%
final,%
notitlepage,%
oneside,%
twocolumn,%
nobibnotes,%
nofootinbib,%
superscriptaddress,%
noshowpacs,%
centertags]%
{revtex4}

\usepackage{multirow}
\renewcommand{\baselinestretch}{0.8}
\begin{document}

\selectlanguage{english}

\title{Star Formation in Nearby Isolated Galaxies}

\author{\firstname{I.~D.}~\surname{Karachentsev}}
\affiliation{\saoname}
\affiliation{Leibniz-Institut f\"ur
Astrophysik (AIP), Potsdam, D-14482 Germany}

\author{\firstname{V.~E.}~\surname{Karachentseva}}
\affiliation{Main Astronomical Observatory, National Academy of
Sciences,  Kiev, 03680 Ukraine}

\author{\firstname{O.~V.}~\surname{Melnyk}}
\affiliation{Astronomical Observatory, Taras Shevchenko Kiev National University,
Kiev, 04053 Ukraine} \affiliation{Institut d'Astrophysique et de
G\'eophysique, Universit\'{e} de Li\`{e}ge, B5C Belgique}

\author{\firstname{H.~M.}~\surname{Courtois}}
\affiliation{Universit\'e de Lyon, Institut de Physique
Nucl\'eaire de Lyon, Villeurbanne, 69100 France}

\received{April 23,  2013}  \revised{June 3, 2013}

\begin{abstract}
We use the FUV fluxes measured with the GALEX to study the star
formation properties of galaxies collected in the ``Local Orphan
Galaxies'' catalog (LOG).  Among 517 LOG galaxies having radial
velocities \mbox{$V_{\rm LG} < 3500$}~km/s and Galactic latitudes
\mbox{$|b|
> 15^{\circ}$}, 428 objects have been detected in FUV.  We briefly discuss
some scaling relations between the specific star formation rate
(SSFR) and stellar mass, \mbox{H\,I-mass,} morphology, and surface
brightness of galaxies situated in extremely low density regions
of the Local Supercluster. Our sample is populated with
predominantly late-type, gas-rich objects with the median
morphological type of Sdm. Only 5\% of LOG galaxies are
classified as early types: E, S0, S0/a, however, they
systematically differ from normal E and S0 galaxies by lower
luminosity and  presence of gas and dust. We find that almost all
galaxies in our sample have  their SSFR below 0.4~[Gyr$^{-1}$].
This limit is also true  even for a sample of 260~active
star-burst Markarian galaxies situated in the same volume. The
existence of such a quasi-Eddington limit for galaxies seems to
be a key factor which characterizes the transformation of gas
into stars at the current epoch.
\end{abstract}

\maketitle

\section{INTRODUCTION}

According to current concepts, the transformation of gas into
stars in   galaxies  is controlled by the internal processes and
depends on the mass  and  morphological type of the galaxy.
Furthermore, the global star formation rate is influenced by
external factors: bursts of star formation in  close encounters
or mergers of galaxies, sweeping out of gas from low-mass
companions while they pass through the dense regions of the halo
of the giant (host) galaxy. Another hidden mechanism of evolution
can be the accretion by the galaxy of the warm intergalactic
medium   which presumably holds about 90\% of all baryons in the
Universe~\cite{fuk2004:Karachentsev_n}. The contribution of the
latter factor in the history of star formation remains quite
unclear.

To make clearer the role of internal processes of gas conversion
into stars, we have to explore them in the galaxies, isolated
from their neighbors to the maximum extent. The appearance of the
mass survey of  ultraviolet radiation of galaxies made at the
GALEX space telescope~\cite{gil2003:Karachentsev_n,
gil2007:Karachentsev_n}, opens the potential  for  a detailed
study of star formation rates in nearby isolated galaxies for
which there exist sufficiently detailed data on their structure
and abundance of  gas. Below we consider the features of star
formation in a representative sample of the most isolated
galaxies of the Local Supercluster, based on the data on their
fluxes in the far ultraviolet (FUV) from the
 GALEX satellite. To our knowledge, this work is
the first systematic attempt to analyze the rates of star formation in the
homogeneous sample of single galaxies in the present epoch ($z <
0.01$).

\section{SAMPLE OF ISOLATED NEARBY GALAXIES}

Using the HyperLEDA\footnote{\tt http://leda.univ-lyon1.fr} and
NED\footnote{\tt http://nedwww.ipac.caltech.edu} databases,
Karachentsev, Makarov and Karachentseva compiled a summary of
approximately 11\,000 galaxies of the  Local Universe with  radial
velocities relative to the centroid of the Local Group
\mbox{$V_{{\rm LG}}< 3500$}~km/s at the galactic latitudes
$|b|>15^{\circ}$. Preparing this sample~(11K) we took into account
the new data on radial velocities of galaxies obtained in the
optical and H\,I sky surveys: SDSS, 6dF, HIPASS, ALFALFA.
Furthermore, we have refined or determined for the first time the
morphological types, apparent magnitudes and other parameters for
many galaxies of the 11K-sample.

The application of the new galaxy clustering criterium to the
11K-sample has led to the creation of the catalogs of pairs,
triplets and groups of galaxies in the Local
Universe~\cite{kar2009:Karachentsev_n, mak2008:Karachentsev_n,
mak2011:Karachentsev_n}. Forty-eight percent  of the galaxies that
are usually referred to as ``field  galaxies'' were left outside
of these catalogs. By a consistent application of the two criteria
of isolation  to these galaxies we have compiled a catalog of
520~most isolated objects of the  Local Supercluster and its
surroundings, called the ``Local Orphan Galaxies'' (LOG)
catalog~\cite{kar2011:Karachentsev_n}. The relative number of
isolated galaxies in the LOG (5\%) is about the same as in
the well-known KIG catalog~\cite{kara1973:Karachentsev_n}, the
objects of which have a median velocity of about 5500~km/s.
Judging from the radial velocities and location of nearby
galaxies, the isolated  galaxies of the LOG and KIG catalogs did
not undergo any significant interaction with neighbors during the
past few billions of years and hence their evolution has been
governed by purely internal mechanisms over a long period of time.

We have checked each galaxy of the LOG catalog in the NED database
 for the presence of the FUV ultraviolet flux   in the ($\lambda_{\rm eff}=1539$~\AA, ${\rm FWHM}=269$~\AA)
 band from measurements of the GALEX  orbital telescope ~\cite{gil2003:Karachentsev_n,
gil2007:Karachentsev_n}. In frequent occasions, when the FUV
image of the galaxy was split into several condensations, we
have summed the $F_{\rm FUV}$  flux  throughout the optical disk
of the galaxy.

To determine the global rate of star formation in the galaxy,
SFR, we followed the scheme used in  Lee and et
al.~\cite{lee2011:Karachentsev_n}:
$$
\log({\rm SFR}~[M_{\odot}/{\rm yr}]) = \log F^c_{\rm FUV} + 2\log D - 6.78, \eqno(1)
$$
where $D$ is the distance to the galaxy in Mpc, and the flux
$F_{\rm FUV}$ in mJy{} is corrected for the  extinction of light
$$
\log(F^c_{\rm FUV}/F_{\rm FUV}) = 0.772(A^G_B+A^i_B). \eqno(2)
$$
Here, the value of  Galactic extinction in the $B$-band, $A^G_B$,
was taken according to~\cite{sch1998:Karachentsev_n}, and internal
extinction in the galaxy itself was determined as
$$
A^i_B=(1.54+2.54(\log 2V_m-2.5))\log(a/b) \eqno(3)
$$
 through the apparent galaxy axis ratio  $a/b$ and the amplitude of
internal rotation  $V_m$~\cite{ver2001:Karachentsev_n}.  For the
dwarf galaxies with  \mbox{$V_m<39$ km/s}  and  gas-poor
E,~S0~galaxies, internal extinction was considered negligible.

Supplementing the LOG catalog with the FUV flux values, we have
filled it with new data on the H\,I fluxes of galaxies from
the EDD\footnote{\tt http://edd.ifa.hawaii.edu} database, as well
as checked and refined the data on
 morphological types and apparent
magnitudes of galaxies. Three galaxies have been removed from the
LOG catalog: LOG\,25 (having a new radial velocity of
\mbox{$V_h=5205$ km/s}), LOG\,368 (not completely isolated) and
LOG\,377 (not having a clear optical   identification for the
radio source   HIPASS~\mbox{J\,1615--17}). The updated LOG catalog
is presented in Table~1.

 \renewcommand{\baselinestretch}{1.0}
\begin{turnpage}
 \begin{table*}
\setcaptionmargin{0mm} \onelinecaptionstrue \captionstyle{normal}
 \caption{Parameters of isolated galaxies in the LOG catalog }
\medskip
\begin{tabular}{r|l|c|r|r|l|r|c|c|c|r|c|r|r|r|r|r|c}

\hline
LOG &~~~~~~~Name    &   RA (J2000.0) Dec    &   $D$~~~  &   $B$~~~  &~$A_B^T$   &~$T$ &   SB  &   $\log (a/b)$    &   $K^{c}$ &   $\log {\rm FUV}$    &   $\log F_{\rm H\,I}$ &   $W_{50}$    &   $\log {\rm SFR}$    &   $\log M_*$    &   P~~~    &   F~~~    &   Note    \\
\hline
(1)~~   &~~~~~~~~~~(2)  &   (3) &   (4)~~ & (5)~~~&~(6) &   (7) &   (8) &   (9) &   (10)    &   (11)~~~~~&  (12)    &   (13)    &   (14)~~~~~&  (15)~~~~&   (16)~~& (17)    &   (18)    \\
\hline
1~~   &   ESO149$-$013    &    000246.3$-$524618  &   18.67   &   15.39   &   0.20    &   8   &   N   &   0.40    &   12.59   &   2.70    &   1.06    &   98  &   $-$1.38 &   8.82    &   $-$0.06 &   0.48    &       \\
2~~   &   ESO149$-$018    &    000714.5$-$523712  &   23.89   &   15.78   &   0.13    &   9   &   N   &   0.10    &   13.30   &   2.57    &   0.74    &   103 &   $-$1.35 &   8.75    &   0.04    &   0.35    &       \\
3~~   &   UGC00064    &    000744.0+405232    &   7.59    &   15.5    &   0.36    &   10  &   N   &   0.10    &   12.79   &   2.93    &   1.24    &   60  &   $-$1.82 &   7.96    &   0.37    &   0.31    &       \\
4~~   &   UGC00063    &    000750.8+355759    &   9.79    &   15.34   &   0.27    &   10  &   N   &   0.18    &   12.72   &   2.61    &   0.28    &   42  &   $-$1.98 &   8.21    &   $-$0.04 &   $-$0.27 &       \\
5~~   &   ESO538$-$024    &    001017.8$-$181551  &   19.3    &   15.08   &   0.14    &   8   &   N   &   0.07    &   12.34   &   3.06    &   0.92    &   25  &   $-$1.05 &   8.95    &   0.15    &   0.04    &   *   \\
6~~   &   PGC130903   &    001108.7$-$385915  &   43.56   &   15.36   &   0.07    &   6   &   H   &   0.29    &   12.20   &   2.46    &   0.3:    &   --  &   $-$0.99 &   9.71    &   $-$0.56 &   0.06    &       \\
7~~   &   6dF...  &    001408.3$-$353648  &   44.77   &   16  &   0.25    &   9   &   H   &   0.43    &   13.40   &   2.37    &   0.71    &   117 &   $-$0.92 &   9.25    &   $-$0.03 &   0.42    &       \\
8~~   &   SDSS... &    001500.1$-$110804  &   47.49   &   17.8    &   0.16    &   6   &   N   &   0.65    &   14.54   &   1.56    &   0.3:    &   --  &   $-$1.74 &   8.85    &   $-$0.45 &   0.89    &       \\
9~~   &   ESO241$-$027    &    001502.7$-$431731  &   44.32   &   15.68   &   0.03    &   6   &   H   &   0.16    &   12.55   &   2.57    &   0.05    &   --  &   $-$0.89 &   9.58    &   $-$0.33 &   $-$0.27 &       \\
10~~  &   6dF...  &    001550.9$-$225511  &   44.01   &   15.78   &   0.18    &   6   &   N   &   0.26    &   12.50   &   2.54    &   0.56    &   89  &   $-$0.81 &   9.60    &   $-$0.27 &   0.15    &       \\
11~~  &   ESO194$-$002    &    001830.4$-$473921  &   19.63   &   16.12   &   0.05    &   7   &   L   &   0.09    &   13.22   &   2.39    &   0.14    &   46  &   $-$1.77 &   8.61    &   $-$0.23 &   $-$0.01 &       \\
12~~  &   AM0016$-$575    &    001909.3$-$573830  &   22.41   &   15.36   &   0.18    &   2   &   N   &   0.11    &   11.08   &   2.57    &   1.32    &   141 &   $-$1.37 &   9.58    &   $-$0.81 &   0.88    &   pec \\
13~~  &   UGC00199    &    002051.8+125122    &   27.60   &   17.3    &   0.34    &   8   &   L   &   0.04    &   14.36   &   2.16    &   0.62    &   94  &   $-$1.47 &   8.45    &   0.22    &   0.47    &       \\
14~~  &   ESO150$-$005    &    002225.6$-$533851  &   15.15   &   13.99   &   0.18    &   8   &   N   &   0.15    &   11.21   &   3.31    &   1.14    &   103 &   $-$0.97 &   9.19    &   $-$0.02 &   $-$0.03 &   *   \\
15~~  &   NGC0101 &    002354.6$-$323210  &   46.73   &   13.46   &   0.14    &   6   &   N   &   0.04    &   10.22   &   3.40    &   1.07    &   160 &   0.07    &   10.56   &   $-$0.35 &   $-$0.16 &       \\
16~~  &   UM240   &    002507.4+001846    &   46.53   &   17.5    &   0.10    &   9   &   H   &   0.12    &   15.05   &   1.85    &   0.3:    &   --  &   $-$1.51 &   8.63    &   0.00    &   0.64    &       \\
17~~  &   6dF...  &    002755.3$-$031101  &   46.19   &   15.8    &   0.15    &   6   &   H   &   0.04    &   12.55   &   2.67    &   0.46    &   40  &   $-$0.66 &   9.62    &   $-$0.14 &   $-$0.05 &       \\
18~~  &   UM040   &    002826.6+050016    &   20.86   &   15.3    &   0.13    &   9   &   N   &   0.18    &   12.82   &   2.81    &   0.80    &   91  &   $-$1.23 &   8.82    &   0.09    &   0.16    &       \\
19~~  &   UGC00285    &    002851.1+285622    &   33.26   &   15.55   &   0.38    &   4   &   N   &   0.52    &   11.57   &   2.27    &   $-$0.30 &   106 &   $-$1.18 &   9.73    &   $-$0.76 &   $-$0.59 &       \\
20~~  &   UGC00288    &    002903.6+432554    &   7.68    &   15.64   &   0.33    &   10  &   N   &   0.21    &   12.96   &   2.53    &   0.72    &   45  &   $-$2.22 &   7.90    &   0.02    &   0.20    &   *   \\
21~~  &   UGC00313    &    003126.1+061224    &   30.64   &   14.35   &   0.26    &   7   &   H   &   0.23    &   11.24   &   2.75    &   0.00    &   116 &   $-$0.85 &   9.79    &   $-$0.50 &   $-$0.68 &       \\
22~~  &   HS0029+1748 &    003203.1+180446    &   33.01   &   18.03   &   0.57    &   9   &   H   &   0.56    &   15.11   &   1.92    &   0.3:    &   --  &   $-$1.38 &   8.30    &   0.46    &   0.21    &       \\
23~~  &   ESO294$-$020    &    003209.7$-$401605  &   19.08   &   14.45   &   0.25    &   8   &   N   &   0.12    &   11.60   &   3.10    &   0.44    &   120 &   $-$0.93 &   9.23    &   $-$0.02 &   $-$0.57 &       \\
24~~  &   UGC00328    &    003322.1$-$010717  &   29.30   &   16.2    &   0.27    &   8   &   N   &   0.18    &   13.33   &   3.09    &   1.25    &   137 &   $-$0.54 &   8.91    &   0.68    &   0.22    &       \\
\hline
\end{tabular}
\end{table*}
\end{turnpage}
\renewcommand{\baselinestretch}{0.8}

The table columns contain:
\begin{list}{}{
\setlength\leftmargin{2mm} \setlength\topsep{1mm}
\setlength\parsep{-0.5mm} \setlength\itemsep{2mm} }

\item (1) the number of the galaxy in the LOG catalog;

\item (2) the name of the galaxy in the known catalogs;

\item (3) equatorial coordinates for the epoch J2000.0;

\item (4) distance to the galaxy $D=V_{\rm LG}/H_0$ in Mpc,
determined from the radial velocity relative to
Local Group at the Hubble parameter
$H_0=73$~km\,s$^{-1}$\,Mpc$^{-1}$; the cases of use of individual
distance estimates  presented in the NED database are marked by an
asterisk in the last column;

\item (5) apparent magnitude of the galaxy in the  $B$-band;

\item (6) total Galactic and internal extinction in  the
\mbox{$B$-band};

\item (7) morphological type by  de Vaucouleurs scale;

\item (8) index of the average surface brightness of the galaxy:
H for high, N for normal, L for low;

\item (9) logarithm of the apparent axial ratio;

\item (10)   apparent magnitude of the galaxy in the
\mbox{$K_s$-band}, corrected for  Galactic and internal
extinction: \mbox{$K-K^c=0.085(A^G_B+A^i_B)$}; since the majority
of galaxies in the LOG catalog  relate  to the late types for
which the \mbox{2MASS} sky survey greatly underestimates the
integral IR fluxes, we determined the \mbox{$K_s$-magnitude} from
the $B$-magnitude and the average color index:  \mbox{$\langle
B-K\rangle = 4.10$} for the \mbox{$T<3$} types, \mbox{$\langle
B-K\rangle = 4.60-0.2T$} for the \mbox{$T=3$--$8$} types and
\mbox{$\langle B-K\rangle = 2.35$} for \mbox{$T=9$--$10$}
according to the recommendations
from~\cite{jar2003:Karachentsev_n,kar2005:Karachentsev_n};

\item (11) logarithm of the total FUV flux of galaxies
in~\mbox{[mJy]};

\item (12) logarithm of the flux in H\,I  radio line
in~\mbox{[Jy$ \times $km/s]};

\item (13) the H\,I line width on the level of 50\% from the peak in
\mbox{km/s};

\item (14)  star formation rate in the galaxies   (in the units of solar mass a year), computed from  relation (1) accounting for the ratios (2) and (3);

\item (15) logarithm of stellar mass of the galaxy  (in solar
masses), determined from the integral  \mbox{$K_s$-lumi}\-no\-sity
at \mbox{$\langle M_*/L_K\rangle=1$} and apparent magnitude of the
Sun
\mbox{$M_{K,\odot}=3.28$}~\cite{bel2003:Karachentsev_n,bin1998:Karachentsev_n};

\item (16, 17) dimensionless parameters  ${\rm P}$ (Past) and ${\rm
F}$ (Future) that characterize the evolutionary state of the
galaxy: \setcounter{equation}{3}
\begin{eqnarray}
{\rm P}&=&\log({\rm SFR}\times T_0/L_K),  \\[+5pt]
{\rm F}&=&\log(1.85\times M_{\rm H\,I}/{\rm SFR}\times T_0),
\end{eqnarray}
where $T_0=13.7\times 10^9$~yrs is the age of the Universe,
$M_{\rm H\,I}$ is the hydrogen mass of the galaxy,
\mbox{$M_{\rm H\,I} = 2.356 \times 10^5\times D^2\times F_{\rm
H\,I}$}, and the
  1.85 coefficient takes into account the contribution of helium and molecular hydrogen in
the total mass of gas~\cite{fuk2004:Karachentsev_n};

\item (18)   notes on the existence of peculiarities (pec) in the
structure of the given galaxy;  the asterisk marks the galaxies
with individual estimates of distances from the NED.

\end{list}

The complete  computer-readable version of Table~1  is accessible
from the Strasbourg astronomical Data Center (CDS).

\section{SOME INTEGRAL PARAMETERS OF  LOG GALAXIES}

The main feature of galaxies of the LOG catalog is the abundance
among them of objects of  late morphological types. The median of
distribution of the LOG galaxies by type falls on the    Sdm
($T=8)$ type. Owing to this,
 more than 90\% of the sample is detected in the  H\,I line, over 80\%
of galaxies have their FUV fluxes and, consequently, integral star
formation rate estimates.

The three panels of Fig.~1 show the distribution of isolated
galaxies from our catalog, by the logarithms of stellar mass,
hydrogen mass and star formation rate, respectively. The median
values of stellar mass,  $2.3\times10^9M_{\odot}$, and hydrogen
mass, $1\times10^9M_{\odot}$, show that this sample is dominated
by the galaxies of moderate to low mass, but with a high content
of the gas component. Individual values of  $\log {\rm SFR}$ in
the LOG galaxies are distributed in a wide range from $+0.34$ to
$-3.67$ with a median of $-1.05$.

\begin{figure}[]
\includegraphics[width=\columnwidth,bb=5 10 266 225,clip]{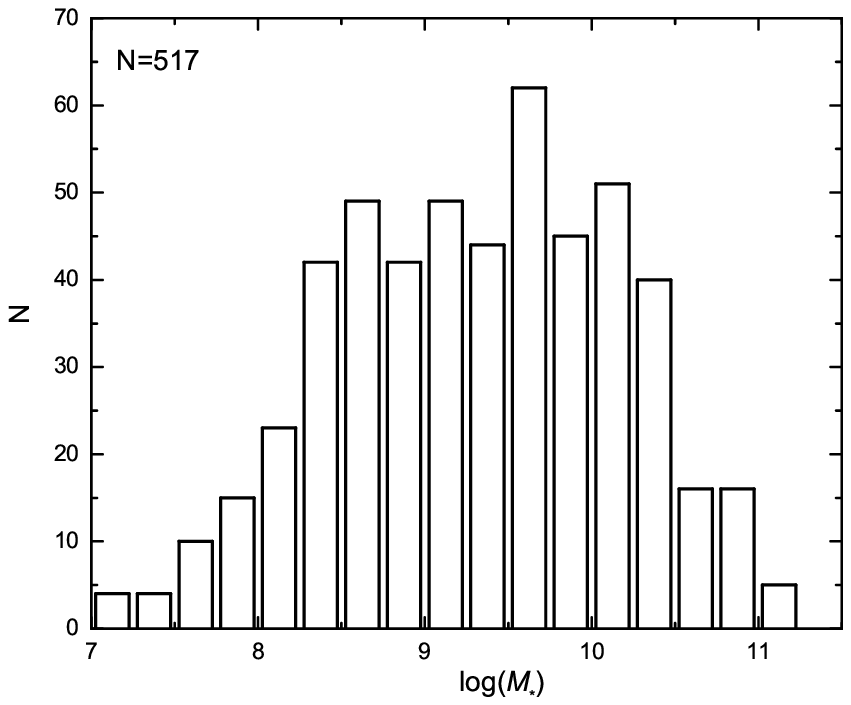}
\includegraphics[width=\columnwidth,bb=5 10 266 225,clip]{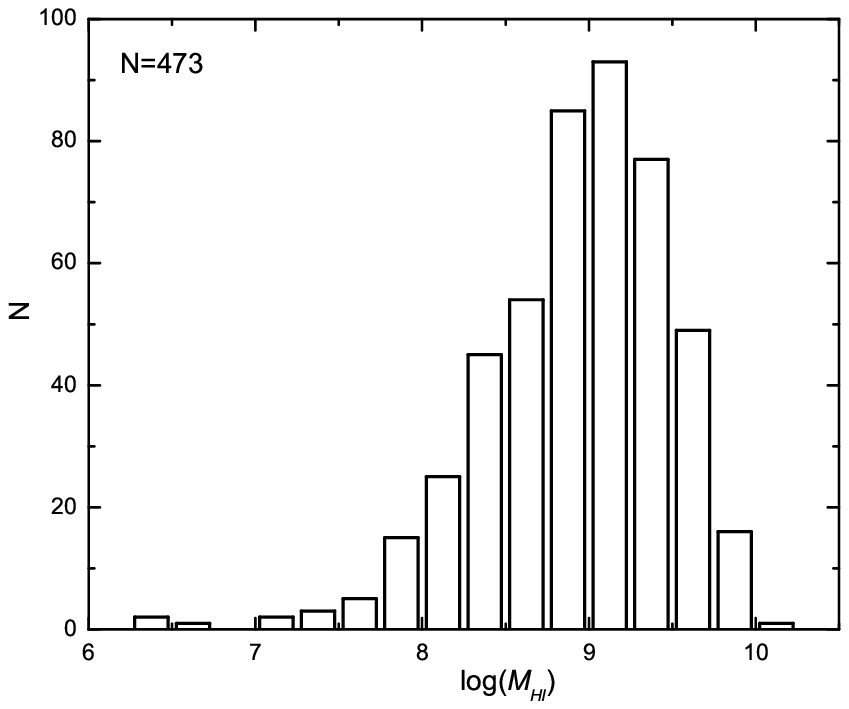}
\includegraphics[width=\columnwidth,bb=5 20 266 235,clip]{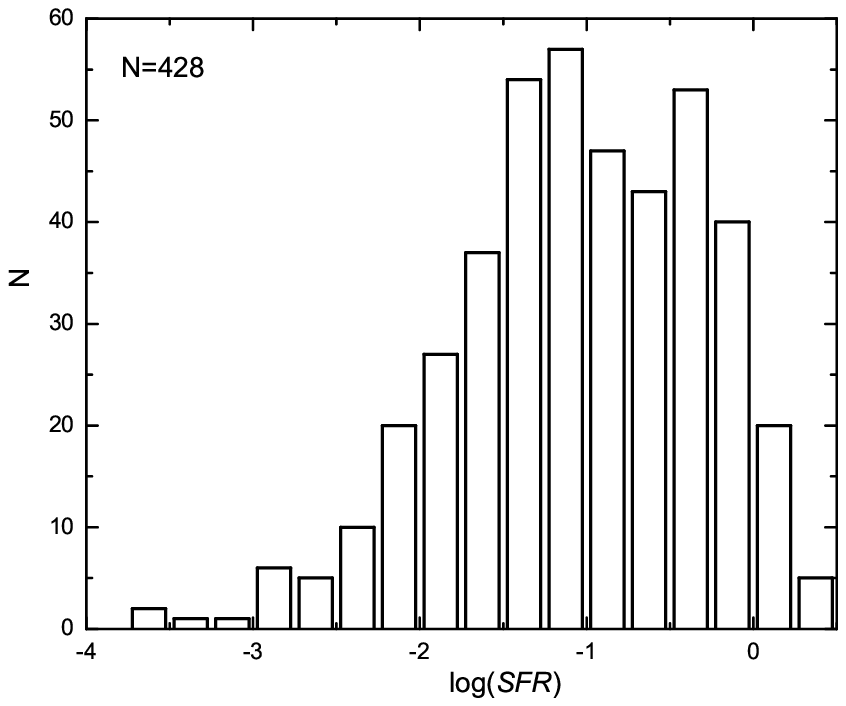}
\captionstyle{normal} \caption{Distribution of isolated galaxies
by the stellar mass (top panel),   hydrogen mass (middle panel)
and integral star formation rate (lower panel).}
\end{figure}

As we can see from Fig.~2, the hydrogen mass-to-stellar mass
ratio increases systematically from the normal luminosity galaxies
to dwarf systems, described by the regression
$$\log(M_{\rm H\,I}/M_*)= -0.54\log(M_*)+4.65 \eqno(6) $$
with a correlation coefficient  $R=-0.76$ and   standard
deviation ${\rm SD} = 0.40$.  In some dwarf galaxies about 90\%
of baryon mass falls to the gas component. Such objects are
obviously in the early stages of the process of transformation of
their gas into stars.

\begin{figure}
\setcaptionmargin{5mm}
\onelinecaptionsfalse
\includegraphics[width=\columnwidth,bb=5 15 266 225,clip]{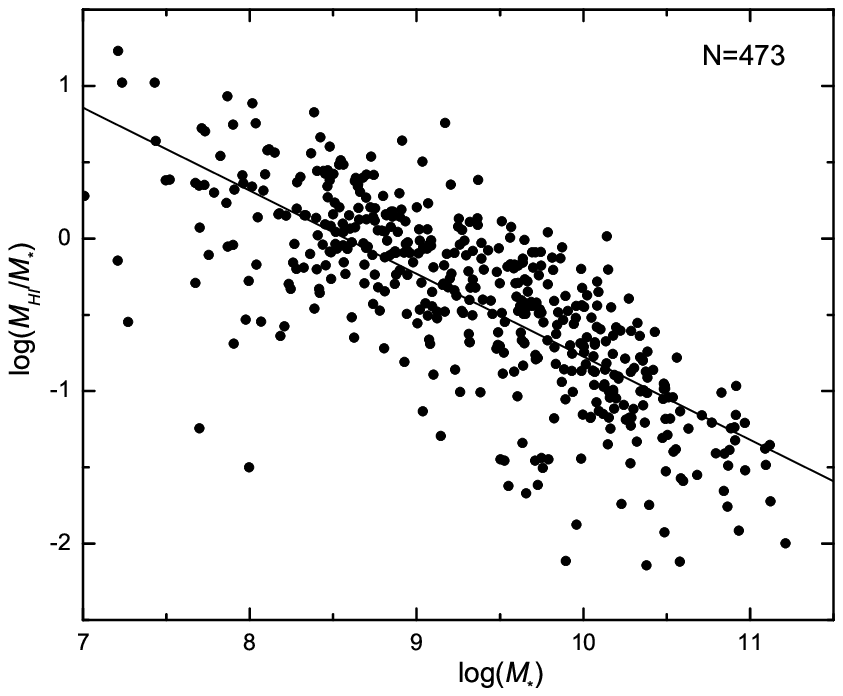}
\captionstyle{normal} \caption{The hydrogen mass-to-stellar mass
ratio for the isolated galaxies of different stellar masses.}
\end{figure}

Figure~3 reproduces the distribution of \mbox{LOG} galaxies by
the value of  integral star formation rate and hydrogen mass. The
solid line in the figure corresponds to the power law $\log {\rm
SFR} \propto 3/2 \log(M_{\rm H\,I})$, which was dubbed the
Kennicutt--Schmidt law~\cite{ken1998:Karachentsev_n}. As we can
see, apart from several objects, the majority of isolated
galaxies follow the established relation quite well, which
looks even  clearer if we exclude the early-type galaxies.

\begin{figure}[tbp!!!]
\setcaptionmargin{5mm}
\onelinecaptionsfalse
\includegraphics[width=\columnwidth,bb=5 15 266 227,clip]{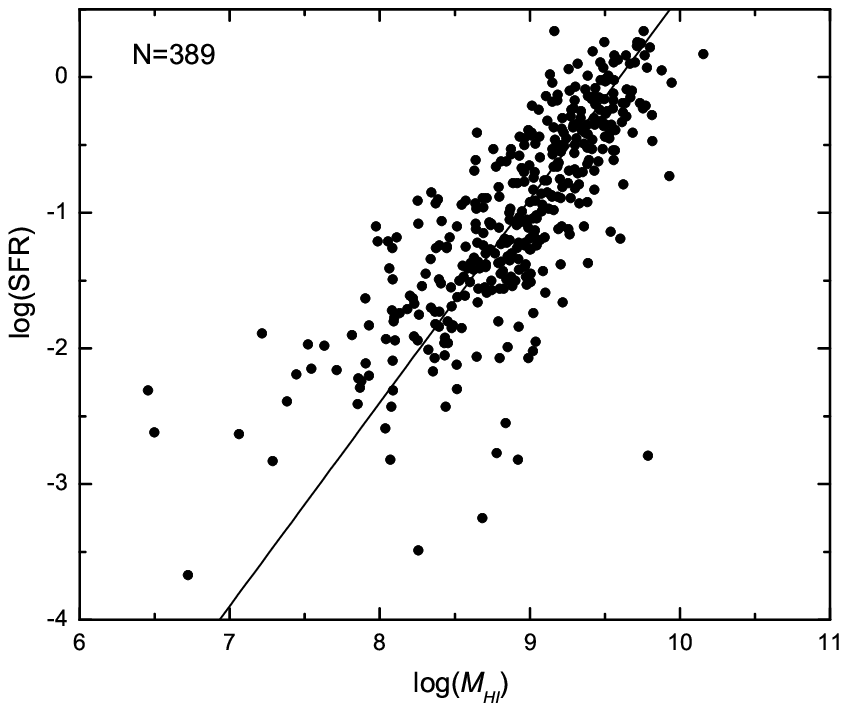}
\captionstyle{normal} \caption{Integral star formation rate in
isolated galaxies of different hydrogen masses. The line
represents the Kennicutt--Schmidt power law with the 3/2
exponent.}
\end{figure}

\section{SPECIFIC  STAR FORMATION RATE AND  GAS RESERVES IN THE GALAXIES}

An important characteristic of a galaxy is the specific rate of
star formation, normalized per unit  of its
\mbox{$L_K$-luminosity} or stellar mass,  ${\rm SSFR}={\rm
SFR}/M_*$. The variation of this value depending on the
stellar mass of the isolated galaxy is presented in Fig.~4. The
left panel of the figure denotes the early-type
(\mbox{$T\leq1)$}, intermediate ($T=2$--$8$) and late-type
($T=9,10$) galaxies  by different symbols. As might be expected,
a limited population of E and S0-galaxies has systematically depressed values of
$\log {\rm SSFR}$ with the median of $-11.5$. The subsystem of
disk galaxies of \mbox{Sab--Sdm} types is characterized by an
order of magnitude larger median value, $-10.3$, and shows a
trend of decreasing average star formation rate with increasing
stellar mass of the galaxy. Low-mass galaxies of the latest
types: Ir, Im, BCD have a median of   \mbox{$\log {\rm SSFR} =
-10.1$}~[yr$^{-1}$], comparable with the value of the Hubble
constant, \mbox{$\log H_0=-10.14$}~[yr$^{-1}$].

\begin{figure*}[tbp!!!]
\setcaptionmargin{5mm} \onelinecaptionsfalse \vspace{2mm}
\includegraphics[width=\columnwidth,bb=5 20 266 230,clip]{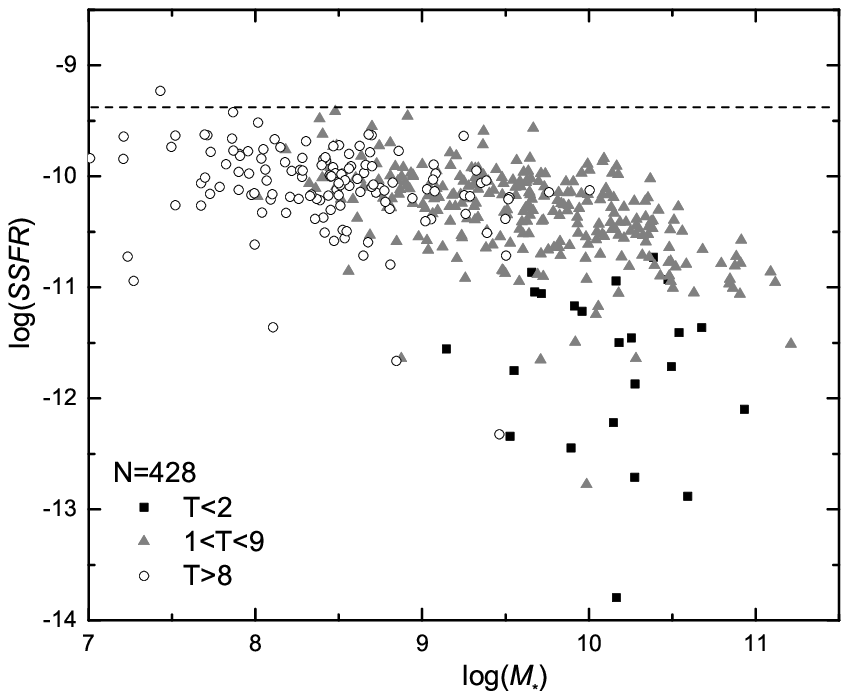}
\includegraphics[width=\columnwidth,bb=5 15 266 225,clip]{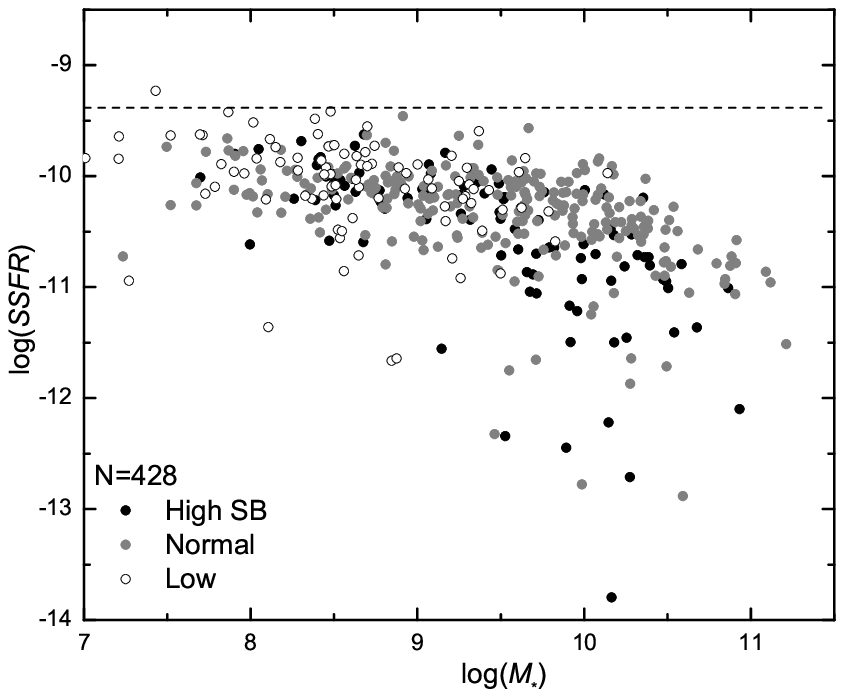}
\captionstyle{normal} \caption{Specific star formation rate and
stellar mass in isolated galaxies of different morphological types
(left panel) and different classes of surface brightness (right
panel). The horizontal line corresponds to the limit \mbox{$\log
{\rm SSFR} = - 9.4$}~[yr$^{-1}$].}
\end{figure*}

The right panel of Fig.~4 presents the same distribution of 428
isolated galaxies by  $\log {\rm SSFR}$ and $\log M_*$,  given
with galaxies marked with the indices of mean surface brightness.
The
 highest SFR with a median of $-10.0$ holds for
the galaxies of low surface brightness, whereas in the galaxies
of normal and high surface brightness  the medians  $\log {\rm
SSFR}$  are $-10.2$ and $-10.4$, respectively.

\begin{figure*}[tbp!!!]
\setcaptionmargin{5mm} \onelinecaptionsfalse \vspace{1mm}
\includegraphics[width=\columnwidth,bb=5 15 266 225,clip]{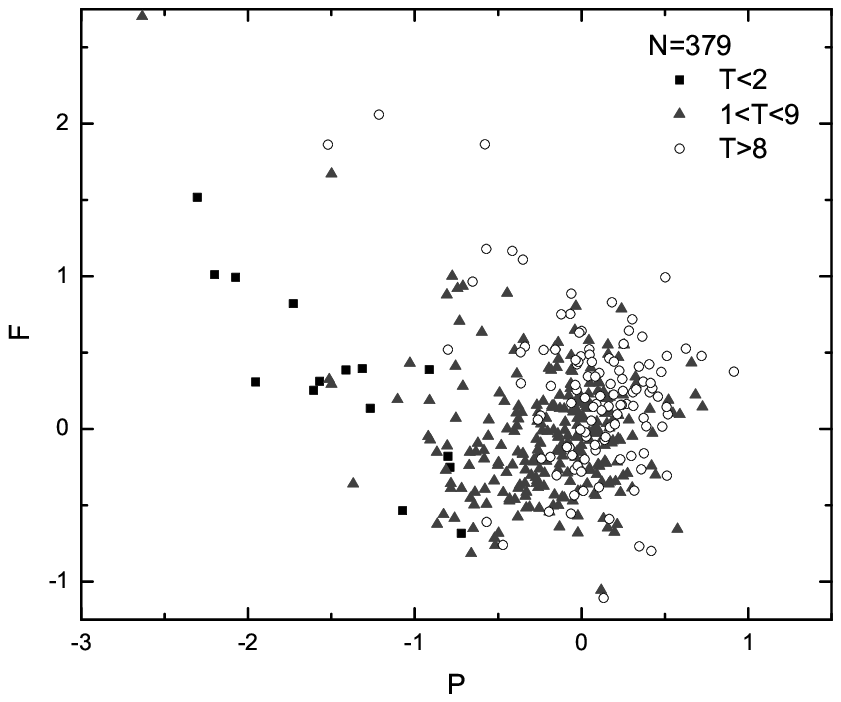}
\includegraphics[width=\columnwidth,bb=5 15 266 225,clip]{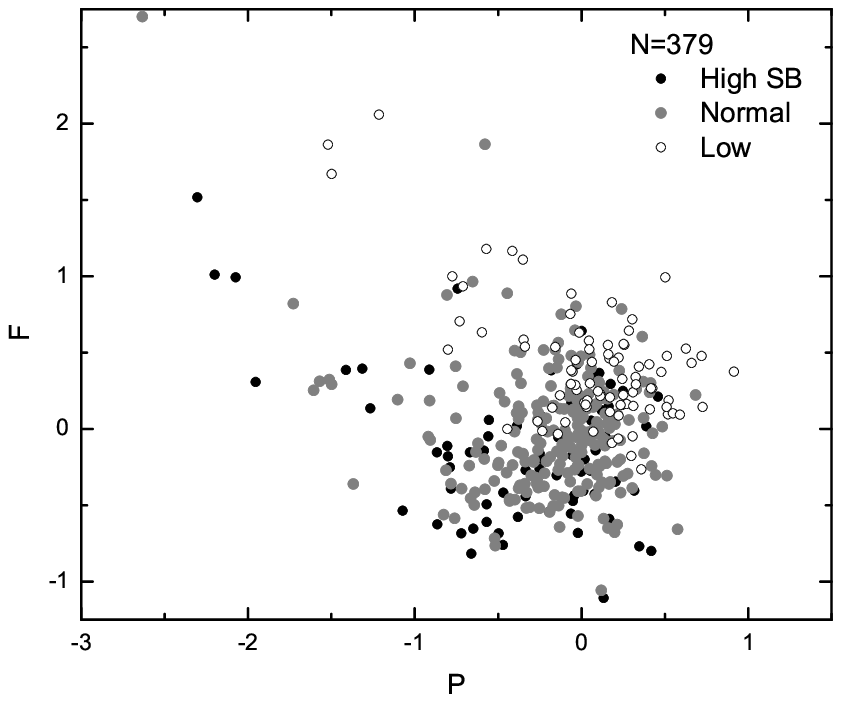}
\captionstyle{normal} \caption{The diagnostic diagram
{Past--Future} of isolated galaxies of different morphological
types (left panel) and different classes of surface brightness
(right panel).} \vspace{1mm}
\end{figure*}

As noted in~\cite{kar2013:Karachentsev_n}, the SSFR in
galaxies of various types of mass and structural types
does not exceed some maximum value of \mbox{$\log {\rm
SSFR}_{\rm max}\simeq -9.4$~[yr$^{-1}$]}. This limit is
indicated in Fig.~4 by a dotted line. Just one isolated galaxy,
\mbox{LOG\,58~$=$ UGCA\,20}, is located above this line. However,
the error in determining its apparent magnitude   is around
$0\fm5$, and in actual fact this irregular galaxy of low surface
brightness can be located below this limit. The presence of the
upper limit in the rates of transformation of gas into stars in
the galaxies is an important parameter of this process, similar to
the Eddington limit for stellar luminosity.

It is convenient to characterize the evolutionary status of
galaxies by  the dimensionless parameters {\rm P} (Past) and {\rm
F} (Future), which are independent of the   galaxy distance
measurement
errors~\cite{kar2007:Karachentsev_n,kar2010:Karachentsev_n}. The
diagnostic diagram ({\rm P}, {\rm F}) for the isolated galaxies
is presented in Fig.~5. On its left panel, the LOG galaxies
are divided into three categories based on morphological types:
(E--Sa), \mbox{(Sab--Sd)} and \mbox{(Im, BCD, Ir)}, while on the
right panel they are sorted by the
 index of average surface brightness: high, normal, low.

According to the relations (4) and (5), the galaxy
located in the center of the diagram  \mbox{(${\rm P}=0$, ${\rm
F}=0$)} is able to reproduce its observed   $L_K$ luminosity
(stellar mass) during the Hubble time at the currently observed
star formation rate; and the gas reserves in it are sufficient to
support the observed SFR on the scale of  yet
another Hubble time.

The median values of the parameters {\rm P} and {\rm F} for the
galaxies of the above categories are listed in Table~2. As
follows from these data, in general the population of isolated
galaxies is concentrated towards the origin  \mbox{(${\rm P}=0$,
${\rm F}=0$)} with the typical spread of  $\sigma({\rm
P})\simeq\sigma({\rm F})\simeq0.6$. This means that on the
average the current star formation rates in isolated galaxies are
in accord with their observed luminosities and their gas reserves
are by now exhausted  only half-way.

\begin{table}[b]
\setcaptionmargin{0mm} \onelinecaptionsfalse \captionstyle{normal}
\caption{Medians of the parameters {\rm P}, and {\rm F} for different samples of
isolated galaxies}
\medskip
\begin{tabular}{l|r|r}
 \hline
\multirow{2}{*}{Galaxy type}& \multicolumn{2}{c}{Median} \\
\cline{2-3}
& \multicolumn{1}{c|}{\rm P}& \multicolumn{1}{c}{\rm F}\\
\hline
$T<2$ & $-$1.41 & 0.31 \\
$T=2$--$8$ & $-$0.11 & $-$0.05\\
$T=9$, $10$& 0.09& 0.22\\
High SB& $-$0.16 & $-$0.15\\
Normal SB &$-$0.08& $-$0.06\\
Low SB& 0.16& 0.33\\
All types & $-$0.05& 0.03\\
\hline
\end{tabular}
\end{table}

The variations of  median values in Table~2 show that over the
past  epochs both the early-type galaxies and high surface
brightness galaxies have had significantly higher star formation
rates than those currently observed. Judging by the trend of the
{\rm F} parameter, the high surface brightness galaxies have
already passed half of their evolutionary path, while the objects
of low surface brightness are still  at the early stage of
transformation of gas they have into stars.

\begin{figure*}[tbp!!!]
\setcaptionmargin{5mm} \onelinecaptionsfalse \vspace{2mm}
\centerline{
\includegraphics[width=0.28\textwidth]{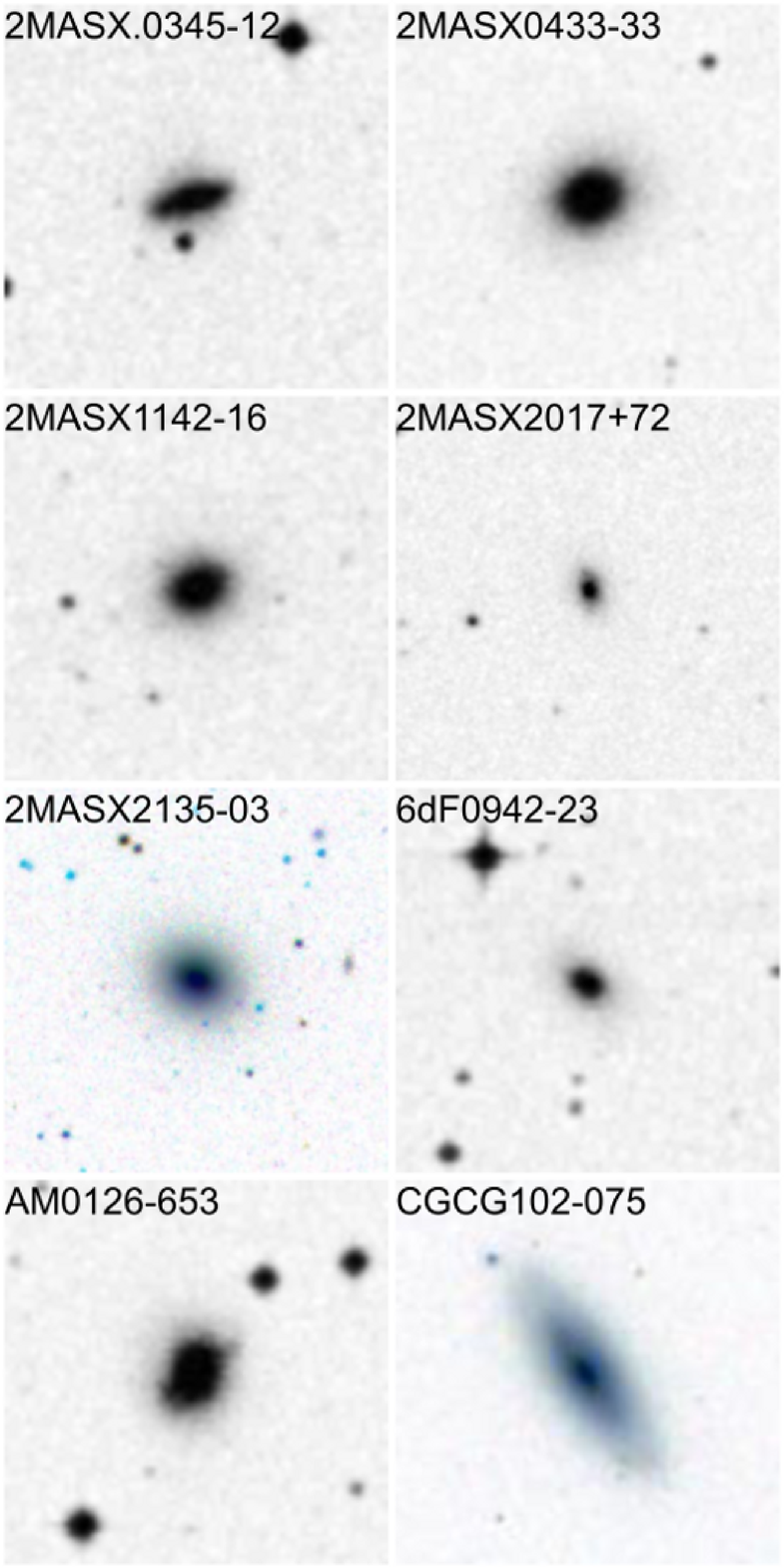}
\hspace{-5pt}
\includegraphics[width=0.28\textwidth]{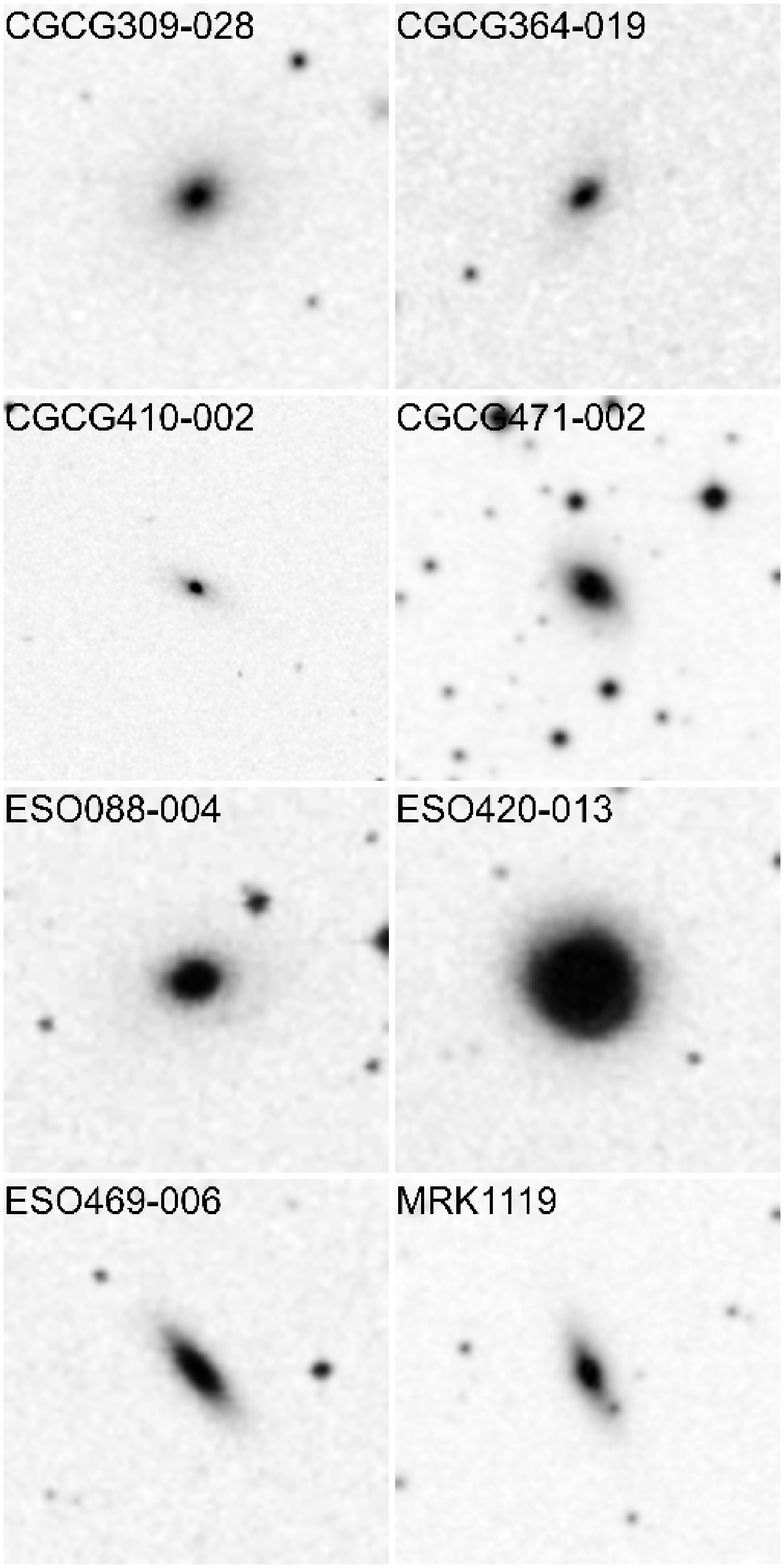}
\hspace{-5pt}
\includegraphics[width=0.28\textwidth]{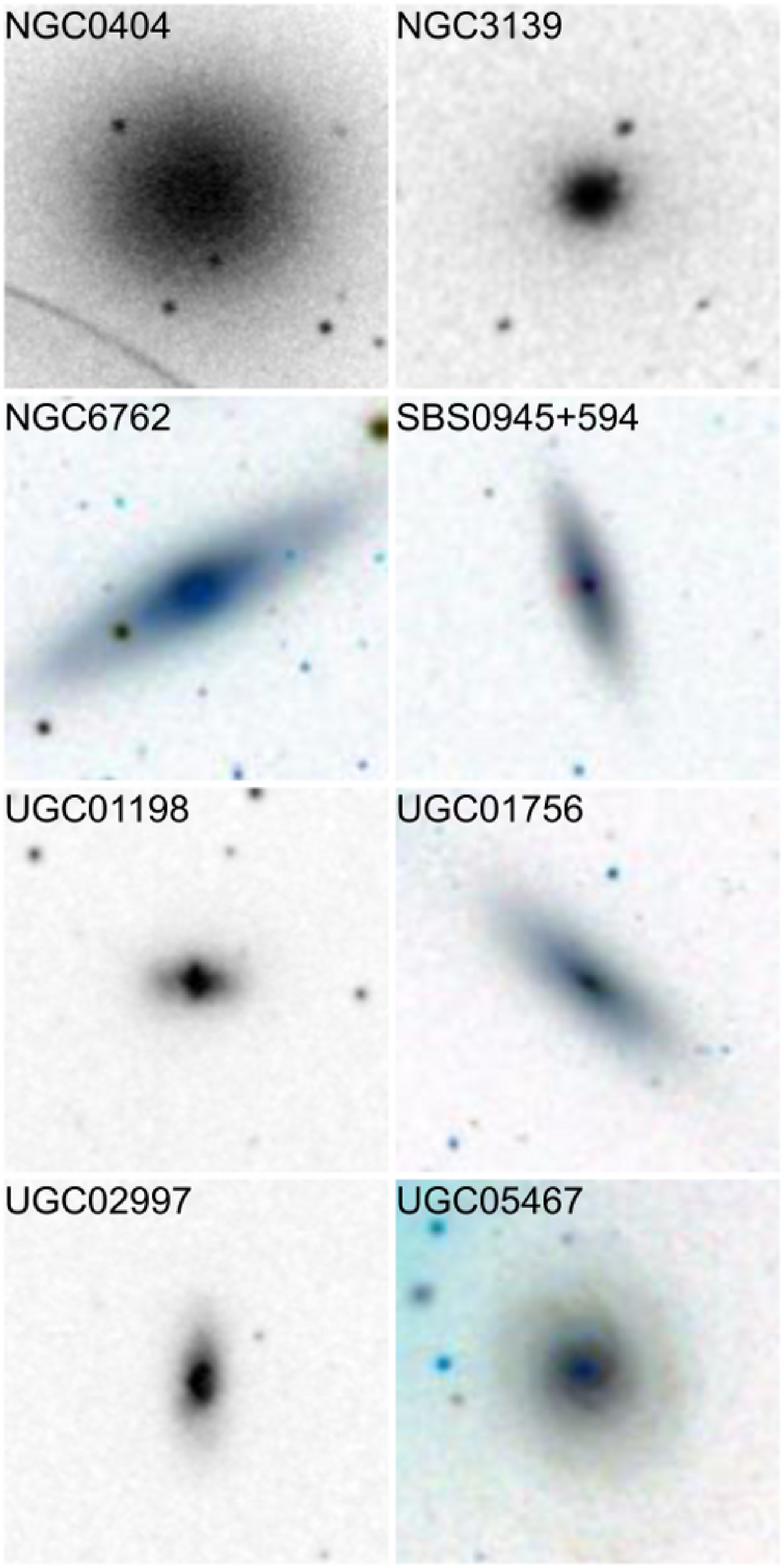}
\hspace{-7pt} \raisebox{208pt}{
\includegraphics[width=0.14\textwidth,bb=14 280 269 537,clip]{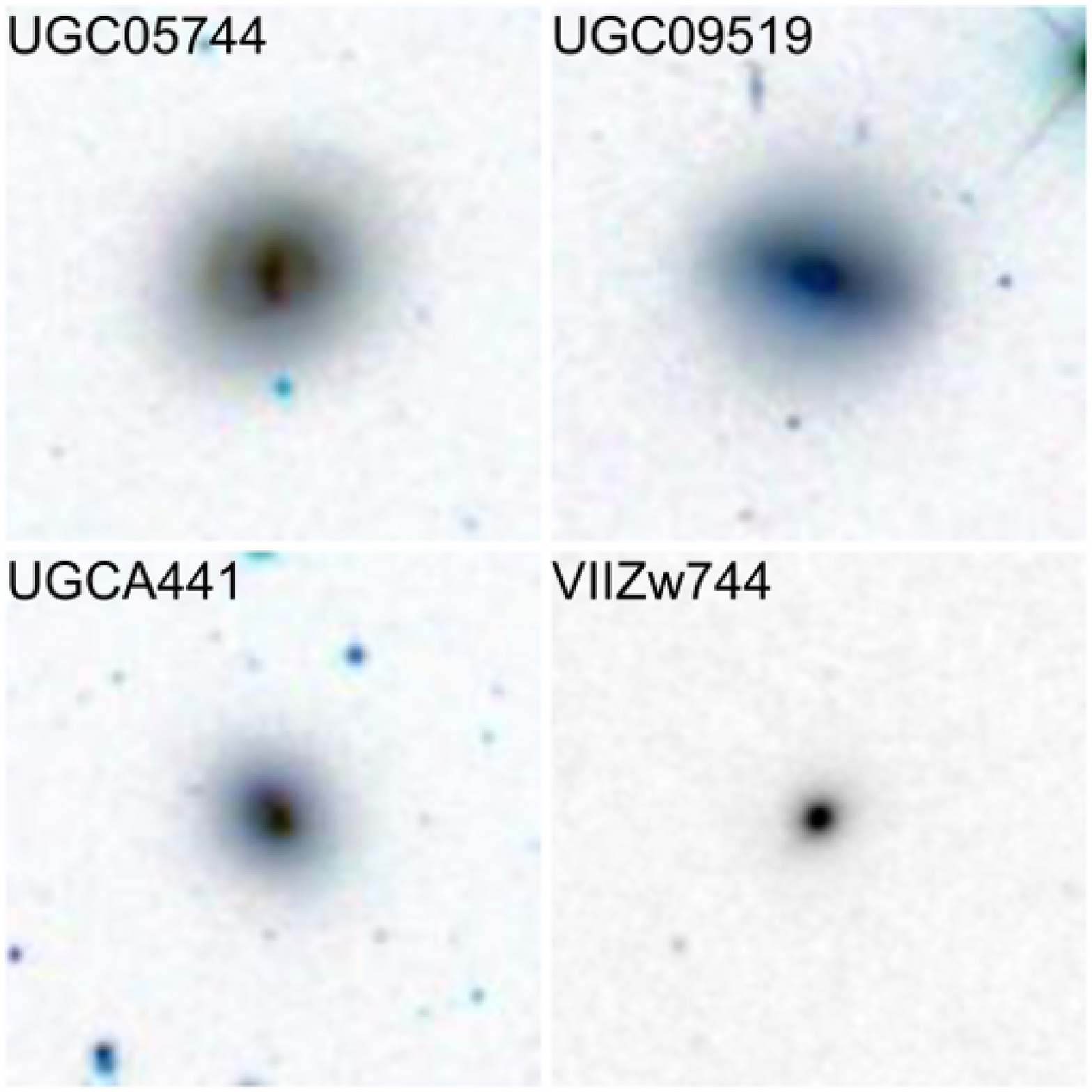}
 }
\hspace{-79pt} \raisebox{139pt}{
\includegraphics[width=0.14\textwidth,bb=270 280 527 537,clip]{Karachentsev_fig6d.eps}
 }
\hspace{-79pt} \raisebox{66pt}{
\includegraphics[width=0.14\textwidth,bb=14 14 269 280,clip]{Karachentsev_fig6d.eps}
 }
\hspace{-79pt} \raisebox{-3pt}{
\includegraphics[width=0.14\textwidth,bb=270 14 527 280,clip]{Karachentsev_fig6d.eps}
 }
} \captionstyle{normal} \caption{Reproductions of images of the
isolated early-type galaxies sized  $2^{\prime}\times 2^{\prime}$
from the  SDSS and POSS-II sky surveys. North is on top, east is on the left. }
\end{figure*}

\begin{table*}[tbp]
\setcaptionmargin{0mm} \onelinecaptionstrue \captionstyle{normal}
\caption{Isolated early-type galaxies}
\medskip
\begin{tabular}{r|l|c|r|r|c|r|r|r|c} \hline
 LOG& \hspace{1.5cm} Name   &  RA(J2000.0)Dec & $D$~~~  & $T$ & SB&  $K_s$~~   & $\log F_{\rm H\,I}$& $\log {\rm FUV}$& Note \\ \hline
31~~  &   CGCG410$-$002   &    004448.4+050809    &   42.00   &   0   &   H   &   10.99   &   0.3:    &   2.67    &   IR  \\
50~~  &   NGC0404 &    010927.0+354304    &   3.05    &   0   &   N   &   6.83    &   1.59    &   3.39    &   IR  \\
54~~  &   AM0126$-$653    &    012822.4$-$651615  &   20.01   &   1   &   H   &   10.50   &   0.95    &   2.64    &   IR  \\
62~~  &   UGC01198    &    014917.7+851538    &   22.20   &   0   &   H   &   10.12   &   0.02    &   2.38    &   IR  \\
70~~  &   UGC01756    &    021653.9+021212    &   42.44   &   0   &   N   &   10.18   &   0.34    &   1.81    &   IR  \\
96~~  &   2MASXJ034559.4$-$123149 &    034559.4$-$123149  &   12.33   &   1   &   H   &   10.87   &   0.3:    &   1.99    &   IR  \\
104~~ &   ESO420$-$013    &    041349.7$-$320025  &   47.25   &   0   &   H   &   9.32    &   0.3:    &   2.30    &   IR  \\
105~~ &   UGC02997    &    041604.9+081049    &   21.78   &   1   &   N   &   8.53    &   0.41    &   --  &   IR  \\
112~~ &   2MASXJ043342.0$-$333046 &    043342.0$-$333046  &   36.67   &   $-$1    &   H   &   10.73   &   0.3:    &   1.53    &   --  \\
164~~ &   ESO088$-$004    &    071006.5$-$631544  &   27.90   &   1   &   H   &   9.29    &   0.3:    &   --  &   IR  \\
169~~ &   CGCG309$-$028   &    071804.4+682034    &   38.30   &   0   &   H   &   10.51   &   --  &   1.03    &   --  \\
220~~ &   6dF...  &    094208.4$-$233544  &   41.68   &   0   &   H   &   12.01   &   $-$0.30 &   --  &   --  \\
221~~ &   SBS0945+594 &    094841.6+591539    &   31.68   &   0   &   N   &   11.05   &   $-$0.59 &   --  &   IR  \\
231~~ &   UGC05467    &    100812.9+184225    &   37.92   &   1   &   H   &   9.99    &   0.64    &   3.00    &   IR  \\
234~~ &   NGC3139 &    101005.2$-$114642  &   15.85   &   $-$2    &   H   &   10.46   &   0.3:    &   1.38    &   --  \\
245~~ &   UGC05744    &    103504.8+463341    &   46.10   &   1   &   H   &   9.91    &   --  &   2.69    &   IR  \\
256~~ &   CGCG364$-$019   &    110734.3+825114    &   25.70   &   0   &   H   &   11.14   &   --  &   2.47    &   --  \\
264~~ &   2MASXJ114234.8$-$165210 &    114234.8$-$165210  &   30.49   &   $-$3    &   H   &   10.97   &   0.5:    &   1.12    &   --  \\
303~~ &   CGCG102$-$075   &    135305.4+155040    &   41.36   &   1   &   N   &   10.67   &   0.5:    &   1.85    &   --  \\
332~~ &   UGC09519    &    144621.1+342214    &   24.41   &   0   &   H   &   9.80    &   --  &   0.3:    &   IR  \\
407~~ &   VIIZw744    &    174137.7+830759    &   29.04   &   $-$1    &   H   &   10.81   &   --  &   2.30    &   --  \\
411~~ &   MRK1119 &    175236.9+374453    &   47.00   &   0   &   H   &   10.68   &   --  &   3.00    &   IR  \\
420~~ &   NGC6762 &    190537.1+635603    &   43.90   &   1   &   N   &   10.01   &   --  &   1.02    &   --  \\
435~~ &   2MASXJ201731.5+720726   &    201731.5+720726    &   36.88   &   $-$1    &   H   &   10.66   &   --  &   0.91    &   IR  \\
450~~ &   CGCG471$-$002   &    211652.9+241215    &   43.34   &   0   &   H   &   10.11   &   0.5:    &   2.03    &   IR  \\
455~~ &   2MASXJ213554.0$-$030853 &    213554.0$-$030853  &   42.10   &   $-$2    &   H   &   10.76   &   0.45:   &   2.19    &   IR  \\
484~~ &   ESO469$-$006    &    225508.0$-$305520  &   41.47   &   0   &   H   &   11.21   &   0.5:    &   --  &   IR  \\
505~~ &   UGCA441 &    233739.6+300746    &   22.74   &   1   &   H   &   10.92   &   $-$0.10 &   2.49    &   IR  \\
\hline
\end{tabular}
\end{table*}

\section{ISOLATED EARLY-TYPE GALAXIES}

After the re-classification of morphological types of the LOG
galaxies done in three years, we found that in 73\% of cases our
independent  type identifications coincide with each other. The
vast majority of the 133 unmatched estimates showed  the
differences of \mbox{$\Delta T=\pm1$}, corresponding to the errors
of the parameters  \mbox{$\Delta\log L_K= \Delta P =\pm0.1$},
which are barely visible on the diagrams of Figs.~4~and~5.

However, there is a small (about 5\%) category
of isolated early-type galaxies,   classifying   which one can easily make a
considerable error. The morphological properties
of these galaxies are often conflicting: a smooth
distribution of light over the disc and red color are sometimes combined with
the presence of   emission in the optical spectrum or in the H\,I line, or
with a significant flux in the ultraviolet, like in the Markarian objects.

A list of 28 such galaxies,  classified by us as E, S0, Sa is presented in Table~3.
The designations of the values  therein are the same as in the
original Table~1; its last column marks the presence in the galaxy of an
infrared  \mbox{IRAS flux}~(IR). The images of these
galaxies  sized   \mbox{$2\arcmin\times2\arcmin$},  taken from the
 SDSS and POSS-II sky surveys are given in the form of a mosaic  in Fig.~6.

This scarce collection of isolated galaxies of types
E~($N=7$), S0~($N=12$) and \mbox{S0--Sa}~($N=9$)   has the
following features. About 79\% of the objects of this subsample
are characterized by  high surface brightness, and the presence
of the FUV flux. Only a quarter of these galaxies is detected in
the H\,I line. About 68\% of the objects are the IRAS sources,
indicating the presence of dust  components in them.  The images
of some galaxies exhibit   low-contrast features of  a spiral
structure (UGC\,5467, UGC\,5744), a polar ring (AM\,0126--653),
or the core emission in  $H\alpha$
(SBS\,0945+594). These features indicate that among the very
isolated galaxies there is a lack of classical  E and
\mbox{S0-galaxies}  with no signs of gas and dust. As it was
pointed out previously~\cite{kar2011:Karachentsev_n},
E~and~\mbox{S0} ``orphan'' galaxies have  systematically lower
luminosities than the members of groups and clusters of the same
types.

The small  population of isolated  E,~\mbox{S0-systems} can be an
important indicator of the process of accretion of warm
intergalactic gas, which is shielded  in the late-type objects by
their own activity of star
formation~\cite{moi2010:Karachentsev_n}. The closest and most
expressive example of this special class of galaxies is NGC\,404,
surrounded by an H\,I-cloud,   the central part  and
  the far periphery of which  have revealed the  regions of
star formation~\cite{del2004:Karachentsev_n,thi2010:Karachentsev_n}.

\section{PECULIAR AND MARKARIAN OBJECTS IN THE LOG CATALOG}

As we have already noted~\cite{kar2011:Karachentsev_n,
kara1973:Karachentsev_n}, the LOG and KIG catalogs of isolated
galaxies contain about 5\% of peculiar objects having noticeable
distortions of the general structure, shape asymmetry or the
presence of tidal ``tails.'' The list of 21 galaxies in the LOG
with the enumeration of their anomalies was shown in Table~3
of~\cite{kar2011:Karachentsev_n}. It has been suggested that
these isolated galaxies could have obtained the peculiarity of
their structure as a result of interaction with dark objects with
masses comparable to the masses of  galaxies themselves. Other
possible explanations for these anomalies of the isolated
galaxies suggest that the observed structural distortions are
caused either by the recent merger of a pair of galaxies, or an
asymmetric starburst on the outskirts of a single galaxy. In
either scenario, a detailed study of the kinematics of such
objects  would help to better understand their nature.

The publication of new releases of the SDSS survey adds to the
list of peculiar isolated galaxies. In this regard, we draw
attention to another two objects:  LOG\,337~$=$
UGC\,9588~$=$ VV\,803 and  LOG\,357~$=$ UGC\,9893~$=$
VV\,720, the reproductions of the  SDSS images of which are
presented in Fig.~7. In the first case, the object appears as a
pair of blue dwarf galaxies with  tails on the stage directly
 before the merger phase.  In the second case, a single blue galaxy
seems to be the result of a recent merger of two dwarf systems
with the formation of a polar ring in the central part. Since the
isolated galaxies reside in  very low-density regions, the cases
of mergers between them should be extremely rare. However, an
example of an interacting triple system in the nearby void has
already been mentioned in the
literature~\cite{bey2013:Karachentsev_n}.

\begin{figure}[tbp!!!]
\setcaptionmargin{5mm} \onelinecaptionsfalse \vspace{1mm}
\includegraphics[width=0.6\linewidth,bb=14 271 272 528,clip]{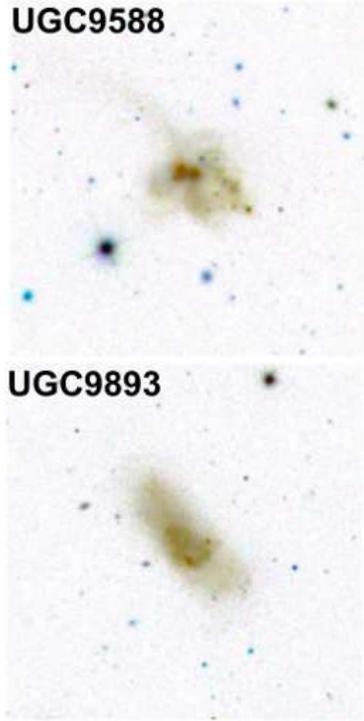}
\includegraphics[width=0.6\linewidth,bb=14 14 272 270,clip]{Karachentsev_fig7.eps}
\captionstyle{normal} \caption{Reproductions of two peculiar
isolated galaxies sized $3^{\prime}\times 3^{\prime}$ from the
SDSS. North is on top, east is on the left.}
\end{figure}

It should be noted that among the 517 galaxies of the LOG catalog
there are 18~active objects from the  Markarian lists. Their total
number in the \mbox{11K-sample} with radial velocities
of \mbox{$V_{\rm LG}<3500$ km/s} is 260, hence their
relative number among the isolated galaxies is not smaller than
among the members of groups and clusters. One would assume that
the star formation activity of Markarian galaxies exceeds the
quasi-Eddington limit of  \mbox{$\log {\rm SSFR}_{\rm
lim}=-9.4$~[yr$^{-1}$]}. Out of 260 Mar\-ka\-rian galaxies located
in the same volume with the LOG objects, 230 have their \mbox{FUV
fluxes} measured. We have estimated their SSFR from these
measurements and compared it with integral luminosity $L_K$. (We
present and discuss these data in a separate paper). As we can see
from the Fig.~8 data, the Markarian galaxies are also located
below the critical value \mbox{$\log {\rm SSFR}=-9.4$}. This fact
reinforces our assertion that the transformation of gas into stars
has a physical limitation by rate, and the dimensionless parameter
\mbox{${\rm dex}({\rm P}_{\rm lim})= T_0\times {\rm SSFR}_{\rm
lim}=5.5$} is an important characteristic of this process.

\begin{figure}[tbp!!!]
\setcaptionmargin{5mm} \onelinecaptionsfalse
\includegraphics[width=0.95\columnwidth,bb=5 15 266 225,clip]{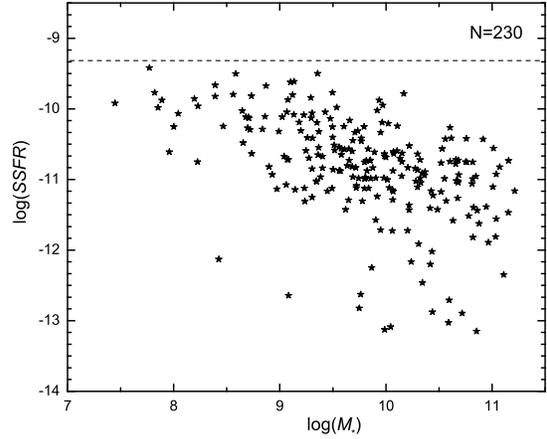}
\captionstyle{normal} \caption{Specific star formation rate and
stellar mass for the  Markarian galaxies in the same volume of the
Local Universe. The horizontal line marks the limit of
\mbox{$\log {\rm SSFR} = - 9.4$}~[yr$^{-1}$]. }
\end{figure}

\section{CONCLUDING REMARKS}

In this paper we continue to study the observational properties of
isolated galaxies located in the nearby Universe within the radius
of about $50$~Mpc. Using   the GALEX  ultraviolet space
telescope data on the FUV fluxes of 389~galaxies from the LOG
catalog, we have estimated their integral star formation rates
SFR. According to our estimates~\cite{kar2011:Karachentsev_n}, the
isolated  LOG galaxies are located in the regions   where the
average local density of matter is about 50~times lower than the
global space density. The LOG sample is dominated by the objects
of the latest types: Sm, Im, BCD, Ir, rich in gas. The transformation
of gas into stars in these isolated galaxies has virtually no
influence of the external factors. The    SFRs of the  LOG
galaxies are not very different from the SFRs of other (non-isolated)
galaxies having the same morphological types. At the same time,
the gas reserves in the LOG galaxies are slightly greater than
those of their non-isolated counterparts.

The specific   star formation rate in the galaxies of different
mass, morphology and surroundings has an upper limit of
\mbox{$\log {\rm SSFR}_{\rm lim}=-9.4$~[yr$^{-1}$]}, which is an
important empirical characteristic of the process of gas
transforming into stars. To our knowledge, the presence of this
quasi-Eddington limit has not yet obtained a direct physical
explanation~\cite{bri2004:Karachentsev_n,hir2013:Karachentsev_n,muz2013:Karachentsev_n,pul2013:Karachentsev_n}.
Although it is quite clear that the rigorous feedback from this
process (gas ejected  by the supernova explosions and radiation
pressure in the conditions of vigorous star formation) should
contribute to setting an upper limit for the SSFR.

About 5\% of the  LOG catalog  objects are classified by us as
elliptical and lenticular galaxies (E,~S0,~S0/Sa). The
very fact of the presence of this category of galaxies among the
particularly isolated objects appears to be a problem, since their
origin involves a series of close encounters and mergers of
galaxies. In fact, the few representatives of isolated E~and S0
galaxies differ from the conventional E~and S0 galaxies in groups
and clusters by their low luminosity and often a  contradictory
combination of a smooth shape,   red color and the presence of
emission lines in the spectrum. If the processes of accretion of
warm interstellar gas significantly affect the increase of the
size and
 mass of the galaxies, the isolated E,~S0~objects can serve
as the most suitable indicators for the study of this process.

\begin{acknowledgments}
This work was supported by the grants of the Russian
Foundation for Basic Research
\mbox{no.~13-02-90407-Ukr-f-a}, 12-02-91338-DFG and the  SFBS of
the Ukraine   F53.2/15. We have made use of the NED ({\tt
nedwww.ipac.caltech.edu}), EDD ({\tt
edd.ifa.hawaii.edu}),  HyperLEDA ({\tt
leda.univ-lyon1.fr}) databases and the data from the Galaxy
Evolution Explorer (GALEX) space observatory.

\end{acknowledgments}

 {}
\end{document}